\documentclass[twocolumn]{aastex631}

\usepackage{subfigure} 
\usepackage{amssymb}
\usepackage{amsmath}
\usepackage{multirow}
\usepackage[para]{threeparttable}
\usepackage{hyperref}

\accepted{for publication in The Astrophysical Journal Supplement}

\shorttitle{Photometric Metallicities for 367,324 stars of $\omega$~Centauri}
\shortauthors{Lu et al.}

\begin{document}

\title{Photometric Metallicities for 367,324 stars of $\omega$~Centauri}

\author[0009-0009-0678-2537]{Xue Lu}
\affiliation{Institute for Frontiers in Astronomy and Astrophysics, Beijing Normal University,  Beijing 102206, China}
\affiliation{School of Physics and Astronomy, Beijing Normal University No.19, Xinjiekouwai St, Haidian District, Beijing, 100875, China}

\author[0000-0003-2471-2363]{Haibo Yuan}
\affiliation{Institute for Frontiers in Astronomy and Astrophysics, Beijing Normal University,  Beijing 102206, China}
\affiliation{School of Physics and Astronomy, Beijing Normal University No.19, Xinjiekouwai St, Haidian District, Beijing, 100875, China}

\author[0000-0002-1259-0517]{Bowen Huang}
\affiliation{Institute for Frontiers in Astronomy and Astrophysics, Beijing Normal University,  Beijing 102206, China}
\affiliation{School of Physics and Astronomy, Beijing Normal University No.19, Xinjiekouwai St, Haidian District, Beijing, 100875, China}

\author[0000-0002-4878-1227]{Tao Wang}
\affiliation{Institute for Frontiers in Astronomy and Astrophysics,
Beijing Normal University, Beijing, 102206, China; }
\affiliation{School of Physics and Astronomy, Beijing Normal University No.19, Xinjiekouwai St, Haidian District, Beijing, 100875, China; }

\author[0000-0003-4573-6233]{Timothy C. Beers}
\affiliation{Department of Physics and Astronomy, University of Notre Dame, Notre Dame, IN 46556, USA}
\affiliation{Joint Institute for Nuclear Astrophysics -- Center for the Evolution of the Elements (JINA-CEE), USA}

\correspondingauthor{Haibo Yuan}
\email{yuanhb@bnu.edu.cn}

\begin{abstract}
$\omega$~Centauri is the most massive and chemically complex multi-population globular cluster with a wide metallicity range that has been extensively studied photometrically and spectroscopically. Using the wide metallicity range of $\omega$~Cen, HST photometry (F275W, F336W, F435W, F625W), and MUSE spectroscopy ([M/H]), we derive [M/H]- and $\rm M_{F625W}$-dependent stellar loci to estimate photometric metallicities from HST colors. Our tests yield metallicity precisions of 0.10\,dex for giants and 0.22\,dex for fainter dwarfs. We construct a photometric metallicity catalog from simultaneous F336W, F435W, and F625W observations (plus F275W where available), containing 20,778 giants and 346,546 dwarfs. A subsample of 20,533 giants is used to study the spatial metallicity distribution and gradient. We find no significant metallicity gradient within the half-light radius, consistent with previous work. Moreover, the previously reported ring-like structure is less pronounced in our data, and no physically significant, irregular two-dimensional metallicity pattern is detected, indicating that the stellar subpopulations are well mixed within the half-light radius. Our catalog significantly extends the metallicity sample of $\omega$~Cen, and this approach can be applied to other HST data to estimate photometric metallicities.

\end{abstract}

\keywords{metallicity; globular star cluster; astronomy data analysis}

\section{Introduction} \label{sec:intro}

Omega Centauri ($\omega$~Cen, NGC 5139) is the optically brightest and most massive known globular cluster in the Milky Way, with an estimated mass of $\sim 3.55 \times 10^{6}\,M_{\odot}$ and a heliocentric distance of only 5.43 kpc \citep{Baumgardt2018,Baumgardt2021}. Due to the pronounced complexity of its stellar populations \citep{Bekki_2003}, it has long been hypothesized that $\omega$~Cen represents the remnant nucleus of a dwarf galaxy that was accreted and tidally stripped by the Milky Way in the distant past. The system exhibits a substantial internal spread in metallicity (e.g., \citealt{Freeman1975,Johnson_2010}) and likely also a non-negligible dispersion in stellar ages \citep{Hilker_2004,Villanova_2014}.

Extensive photometric and spectroscopic datasets are now available, allowing detailed investigations of $\omega$~Cen and, in particular, its multiple stellar populations. Recently, \citeauthor{omegacat1} (\citeyear{omegacat1}, hereafter oMEGACat.I) presented a comprehensive VLT/MUSE dataset extending to the half-light radius of $\omega$~Cen, and provided spectroscopic measurements for several hundred thousand individual stars, constituting by far the largest spectroscopic sample obtained to date for $\omega$~Cen or for any globular cluster. In a companion study, \citeauthor{omegacat2} (\citeyear{omegacat2}, hereafter oMEGACat.II) published a catalog of $\omega$~Cen’s central regions, covering the full half-light radius, based on deep six-band HST photometry (including ultraviolet filters) and high-precision astrometry.

The broad metallicity range of $\omega$~Cen, combined with extensive MUSE spectroscopy from oMEGACat.I and HST UV photometry from oMEGACat.II, allows us to derive photometric metallicity estimates from HST data and thus greatly expand the metallicity sample for $\omega$~Cen. Using photometric methods provides a more complete metallicity census and reduces the target-selection biases associated with spectroscopic studies. 

Compared to spectroscopy, photometric surveys are far more cost effective, reach much fainter magnitudes, and suffer weaker selection effects, making them efficient for inferring stellar parameters in very large samples. Over the past two decades, the rapid growth of wide-field photometric and spectroscopic surveys has made photometric metallicity estimation widely used, especially via stellar-locus fitting methods. Because bluer, narrower filters are generally more sensitive to metallicity, several studies have exploited this: \citet{yuan2015metal} used the $u$-band of the Sloan Digital Sky Survey (SDSS; \citealt{York_2000}), \citet{Huang2022} used the $u/v$-bands of the SkyMapper Southern Sky Survey (SMSS DR2; \citealt{SkyMapper_DR2}), and \citet{Lu_2024} used the bluer NUV-band of the Galaxy Evolution Explorer (GALEX; \citealt{Bianchi2011}), one of the most metallicity-sensitive bands available. However, \citet{Lu_2024} showed that the strong dependence of the NUV-band on surface gravity (or absolute magnitude) must be included in the modeling.

In this study, we combine MUSE spectroscopy from oMEGACat.I with HST photometry from oMEGACat.II to model photometric metallicities. Utilizing the resulting catalog, we subsequently analyze the spatial distribution of metallicity within the cluster and quantify its radial metallicity gradients.
This paper is organized as follows. Section~\ref{sec:data} introduces the data sets. Section~\ref{sec:method} describes our method for estimating photometric metallicity. Section~\ref{sec:result} presents the fitting and test results for the reference sample. Section~\ref{sec:fianl_cat} describes the final metallicity catalog for $\omega$~Cen. Section~\ref{sec:metal_dist} investigates the spatial distribution and radial gradients of metallicty. Section\,\ref{sec:summary} provides a summary.

\section{Data} \label{sec:data}

\subsection{Spectroscopic Data} \label{subsec:spec_data}
We employed data from oMEGACat.I, where the interested reader can obtain a detailed description of the spectroscopic data used.  The spectra were acquired with the Multi-Unit Spectroscopic Explorer (MUSE; \citep{MUSE2010,MUSE2014}), a second generation Very Large Telescope instrument mounted on the UT4 at the Paranal Observatory in Chile and observing in the optical domain (480–930\,nm). It includes both the main observing run (105.20CG.001) and complementary MUSE guaranteed time observations (GTO data). These data together have full coverage out to the half-light radius (4.65' or 7.04 pc, \citealt{Baumgardt2018}) of the cluster, with 342,797 stars. Spectra were extracted for individual stars using PAMPELMUSE \footnote {\href{https://pampelmuse.readthedocs.io/en/latest/about.html}{https://pampelmuse.readthedocs.io/en/latest/about.html}}(\citealt{Kamann2013}) and analyzed with SPEXXY \footnote {\href{https://github.com/thusser/spexxy}{https://github.com/thusser/spexxy}} \citep{Husser2016} to determine key physical parameters, including effective temperature, metallicity, and radial velocity. Extensive error analysis and membership determination were also performed to ensure the robustness of the results.

\subsection{Photometry Data} \label{subsec:photo_data}
We used the HST-based stellar catalog oMEGACat.II, which provides high-precision photometry and proper motions for $\sim$1.4 million stars. The catalog combines over two decades of ACS/WFC and WFC3/UVIS observations from various calibration programs and the dedicated GO-16777 program (PI: A. Seth). It covers a broad wavelength range from the ultraviolet (F275W, F336W) through the optical (F435W, F625W, F658N, F606W) to the near-infrared (F814W), with six-filter coverage across the full field and additional multi-epoch F606W data in the central region. Photometry was obtained with optimized PSF-fitting for crowded fields (see \citealt{Bellini2017}), and proper motions follow the methods of \citet{Bellini2014,Bellini2018} and \citet{Libralato2018,Libralato2022}.  An empirical correction was derived for spatially dependent photometric variations, which considers the spatial dispersion caused by differential reddening and instrumental effects, as described in Section 6.5 of oMEGACat.II.

To combine oMEGACat.II photometry with oMEGACat.I [M/H] values, the catalogs were astrometrically cross-matched using a 40 mas radius ($\sim$1 WFC3/UVIS pixel) and requiring F625W and F435W magnitude differences of $\Delta mag < 0.1$ relative to the \citet{Anderson2010} photometry used for spectral extraction. This yields 307,030 matches out of 342,797 oMEGACat.I stars. In this work, we use the F275W, F336W, F435W, and F625W filters, chosen for their metallicity sensitivity and coverage of $\omega\,Cen$.

\subsection{Construction of the Reference Sample} 
\label{subsec:train_sample}
Based on the 307,030 matched stars, we select a high-quality sample using:

\begin{enumerate}

\item[1)] Spectroscopic label $\rm Flag=1$;

\item[2)] Metallicity errors $e_{MH}<0.1$ for giants and $e_{MH}<0.2$ for dwarfs;

\item[3)] Photometric label $\rm phot\_hq\_flag=1$;

\item[4)] Magnitude errors $e_{mag}\leq0.05$ in F275W, F336W, F435W, and F625W;

\item[5)] Color cuts $\rm 0.7<(F435W-F625W)_0<2.2$ for giants and $\rm 0.55<(F435W-F625W)_0<1.3$ for dwarfs.

\end{enumerate}

We also retain all sources with the brightlist\_flag set to keep the model valid at the bright end. The intrinsic color is
$\rm (F435W-F625W)_0 = F435W-F625W - E(B-V)\times (R_{F435W} - R_{F625W})$,
with the foreground extinction $\rm E(B-V) = 0.185$ from \citet{Clontz2025}. The extinction coefficients for the HST filters are \citet{Schlafly2011}: $\rm R_{F275W} = 5.487$, $\rm R_{F336W} = 4.453$, $\rm R_{F435W} = 3.610$, and $\rm R_{F625W} = 2.219$. Unless noted otherwise, all colors are intrinsic (de-reddened).

After these cuts, the final sample contains 94,276 sources—16,270 giants and 78,006 dwarfs—as shown in Figure~\ref{fig:training_sample}, where the black line marks the adopted giant–dwarf boundary.

\begin{figure}
\centering
\includegraphics[width=\linewidth]{ }
\caption{Color–magnitude diagram of the reference sample, with points color-coded by $\rm [M/H]_{spec}$. The black line follows $F625W = -(F435W - F625W)^2 + 4(F435W - F625W) + 14.8.$ }
\label{fig:training_sample}.
\end{figure}

\section{[M/H]- and $\rm M_{F625W}$-dependent stellar loci} \label{sec:method}
To capture the combined effects of [M/H] and log\textit{g} (here represented by the absolute F625W magnitude, $\rm M_{F625W}$), we fit the stellar loci with a three-dimensional polynomial. The distance adopted for $\omega$~Cen to convert observed magnitudes into absolute magnitudes is 5.426\,kpc (\citealt{Baumgardt2021}). We model the metallicity-sensitive colors F275W$-$F625W and F336W$-$F625W as functions of F435W$-$F625W, [M/H], and $\rm M_{F625W}$.
The fitting equation is expressed as: 
\begin{eqnarray} \label{fit_equation}
f(X, Y, Z) &=& (Y - k)^5 (p_0 X^4 + p_1 X^3 + p_2 X^2 + p_3 X + p_4) \nonumber \\
&& + (p_5 X^4 + p_6 X^3 + p_7 X^2 + p_8 X) \nonumber \\
&& + (p_9 Z^4 + p_{10} Z^3 + p_{11} Z^2 + p_{12} Z) \nonumber \\
&& + p_{13} X^2 Z + p_{14} X Z^2 + p_{15} X Z \nonumber \\
&& + (Y - k)^5 (p_{16} Z^2 + p_{17} Z) + p_{18}, \nonumber \\
\end{eqnarray}
where X, Y, Z represent the color F435W$-$F625W, [M/H], and $\rm M_{F625W}$ and $k=-5.5$. We model the relation as a multivariate polynomial expansion, including 4th-order terms in F435W$-$F625W and $\rm M_{F625W}$, a 5th-order dependence on [M/H] through $(Y-k)^5$, and pairwise interaction (cross) terms. Term powers are chosen empirically after a number of tests.

Note that the color of a star exhibits weaker sensitivity to metallicity when the star is more metal-poor, eventually making it impossible to distinguish between different metallicities at the metal-poor end with a given photometric precision. Therefore following \citet{Huang_bw2025}, [M/H] appears through the factor $(Y-k)^5$, which allows the metallicity sensitivity to diminish toward lower [M/H] and eventually reaches zero at $[M/H]=k=-5.5$.

During fitting, a $3\sigma$-clipping procedure removes outliers. 

We then use a $\chi^2$ method to derive the metallicity, defined as:
\begin{eqnarray}\label{chi2_2}
&&\chi^2(\rm [M/H]) = \nonumber \\ 
&&\sum \limits _{i=1} ^{2} \frac{[\rm C_{i} - C^{model}_{i}(F435W-F625W, [M/H], M_{F625W})]^2}{(\sigma_{c}^i)^2},\nonumber \\ 
\end{eqnarray}
where $\rm C_{i}$ is the intrinsic color observed ($i = 1, 2$ for F275W$-$F625W and F336W$-$F625W, respectively). $\rm C^{model}_{i}(F435W-F625W, [M/H], M_{F625W})$ is the intrinsic color of the model from the stellar loci dependent on [M/H] and $\rm M_{F625W}$, and $\sigma_{c}^i$ are the uncertainties in $\rm C^{obs}_{i}$, respectively.

\section{Result} \label{sec:result}

\subsection{Fitting Results} \label{subsec:fit_result}

\begin{figure*}
    \centering
    \includegraphics[width=\textwidth]{ 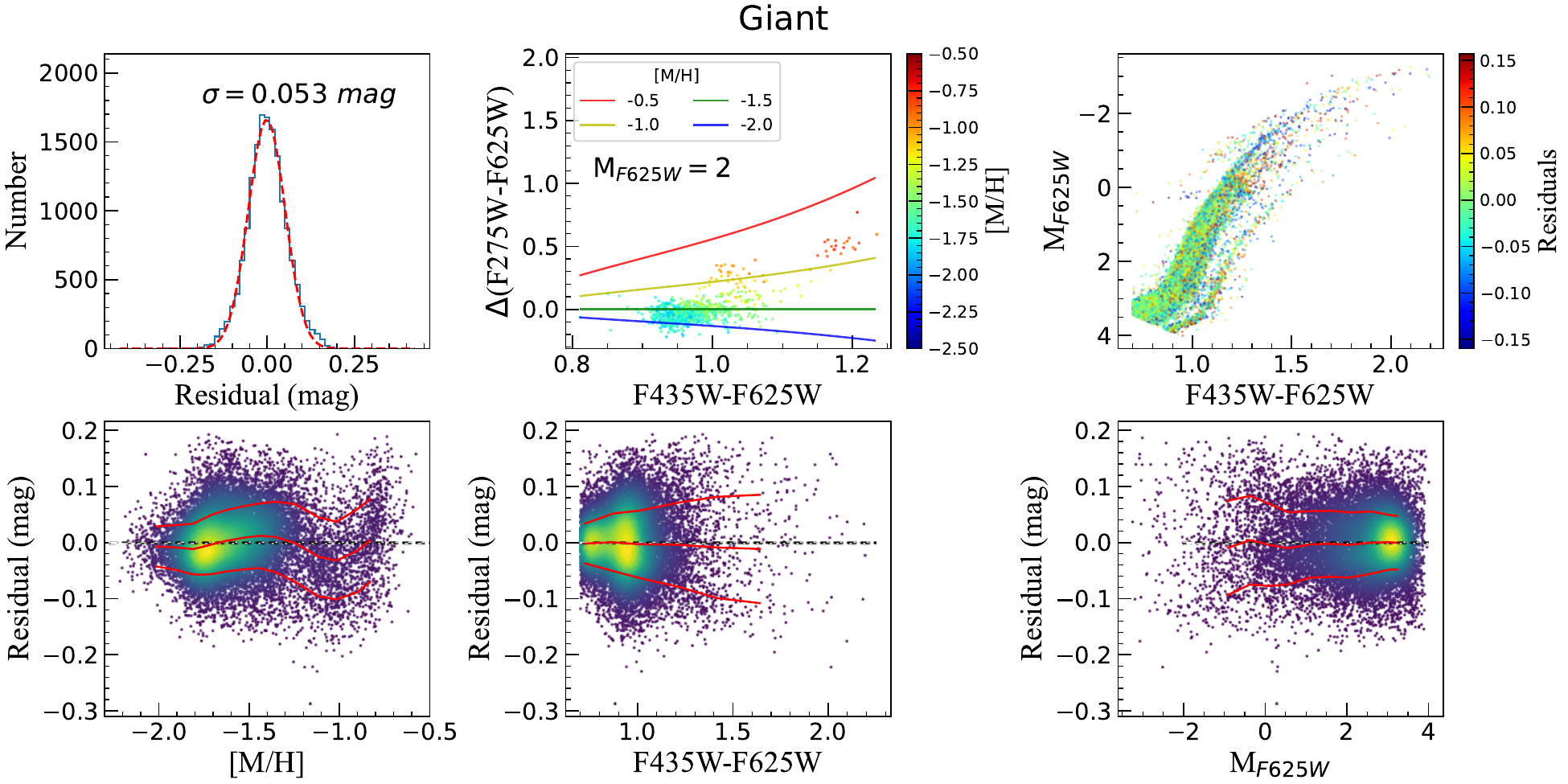}
    \caption{ Fitting residuals of the F275W$-$F625W color. Top left: Histogram of the residuals with a Gaussian fit over-plotted as a dashed-red line. Top iddle: Stellar loci at $\rm M_{F625W}=2$ for different metallicities relative to $\rm [M/H]=-1.5$, with sources satisfying $\rm |M_{F625W}-2|\le0.1$ shown as colored points. Top right: Residuals in the color-magnitude diagram. Bottom panels: Residuals as functions of [M/H], F435W$-$F625W color, and $\rm M_{F625W}$, respectively. Points are color-coded by number density..
    }
    \label{Fig:fit275_giant}
\end{figure*}

\begin{figure*}
    \centering
    \includegraphics[width=\textwidth]{ 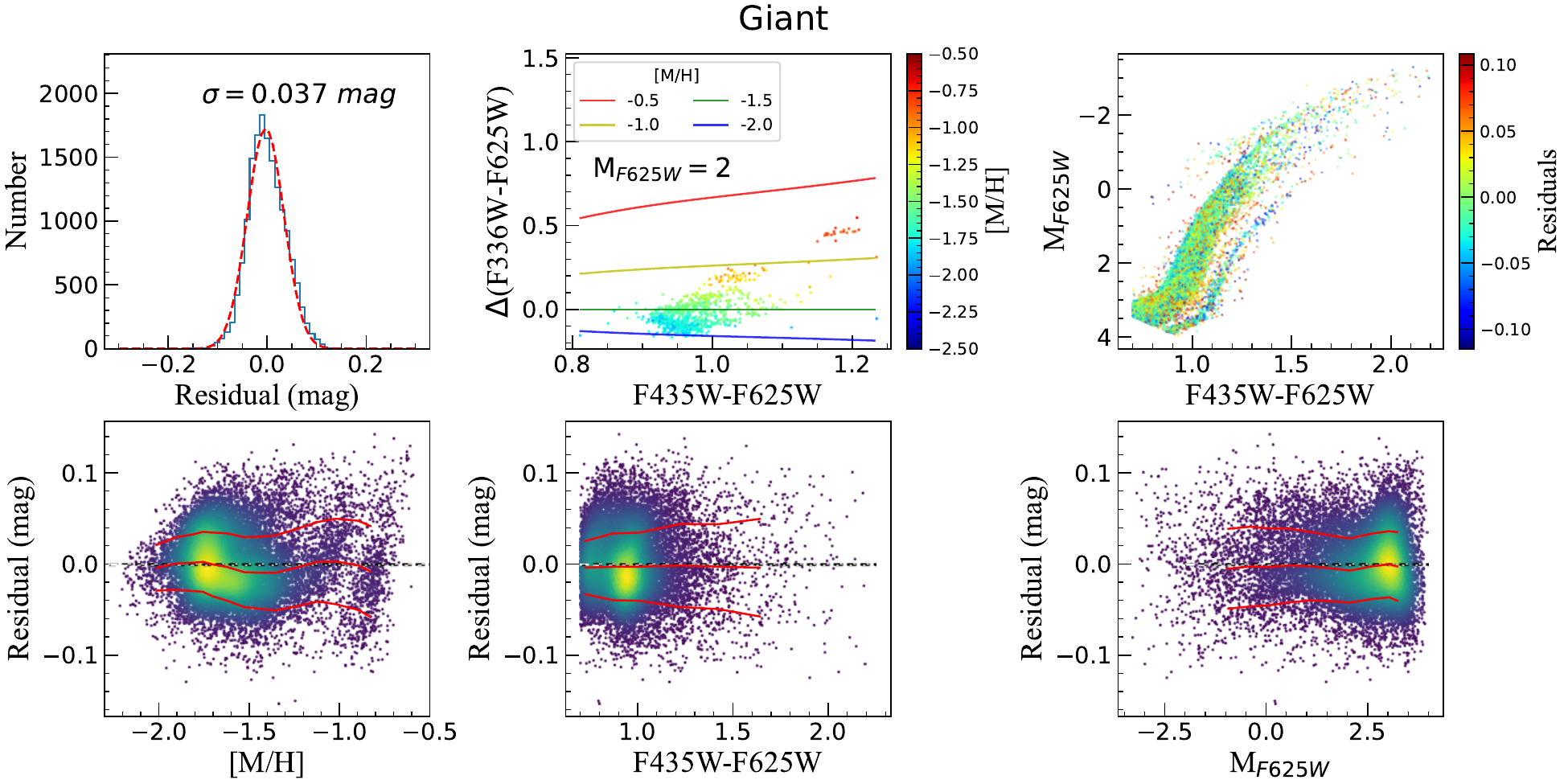}
    \caption{Same as Figure\,\ref{Fig:fit275_giant}, but for the F336W$-$F625W color residuals.}
    \label{Fig:fit336_giant}
\end{figure*}

\begin{figure*}
    \centering
    \includegraphics[width=\textwidth]{ 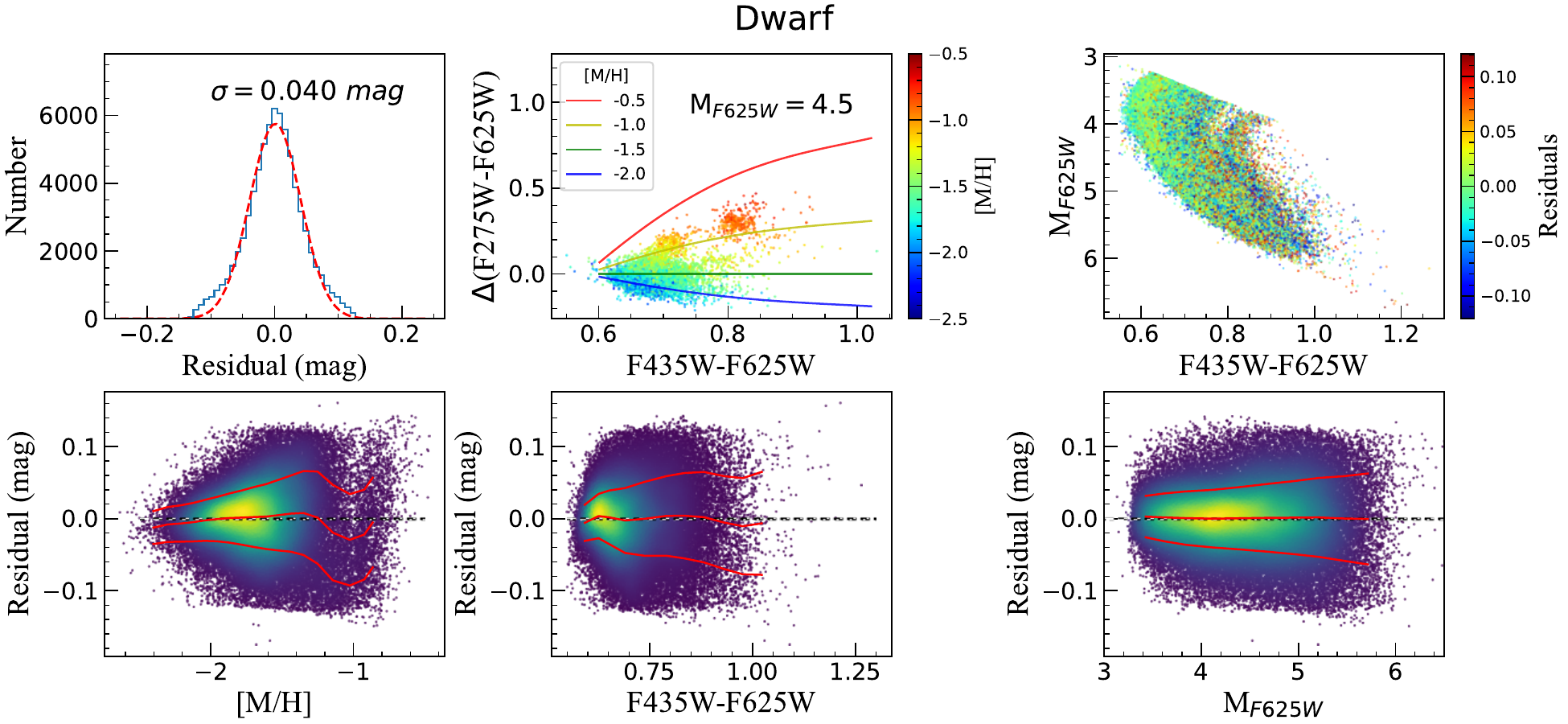}
    \caption{Same as Figure\,\ref{Fig:fit275_giant}, but for dwarfs.}
    \label{Fig:fit275_dwarf}
\end{figure*}

\begin{figure*}
    \centering
    \includegraphics[width=\textwidth]{ 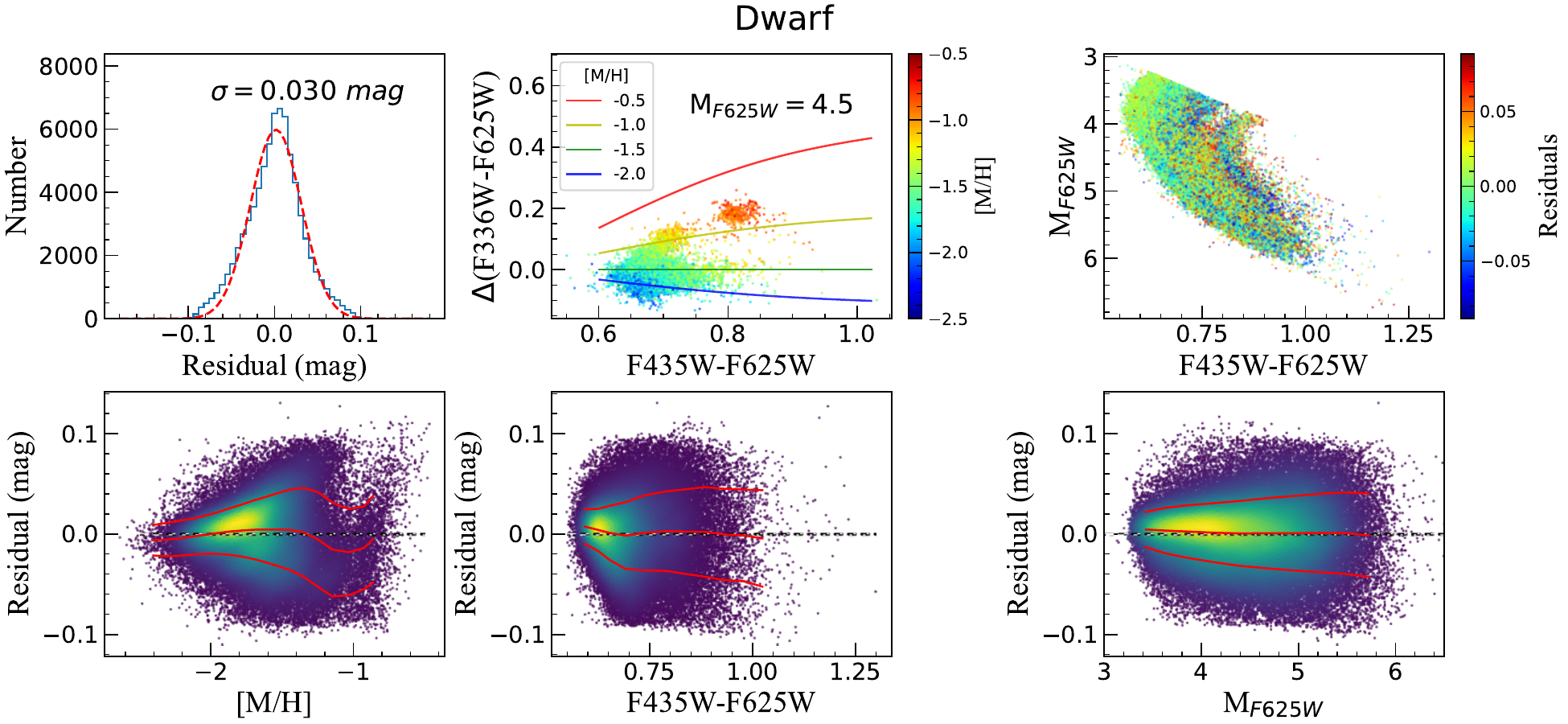}
    \caption{Same as Figure\,\ref{Fig:fit336_giant}, but for dwarfs.}
    \label{Fig:fit336_dwarf}
\end{figure*}

The fitting coefficients $p_i$ for giants and dwarfs are listed in Table\,\ref{tab:fit_coeff}. For the giant reference sample, the fits for F275W$-$F625W and F336W$-$F625W are shown in Figures\,\ref{Fig:fit275_giant} and \ref{Fig:fit336_giant}; the mean residuals are near 0.0 mag, with standard deviations of 0.053 and 0.037 mag, respectively, as indicated in the histograms. For the dwarf reference sample (Figures\,\ref{Fig:fit275_dwarf} and \ref{Fig:fit336_dwarf}), the mean residuals are also near 0.0 mag, with standard deviations of 0.040 and 0.030 mag for F275W$-$F625W and F336W$-$F625W, respectively. In all cases, the residuals exhibit almost no trends with F435W$-$F625W or $\rm M_{F625W}$, and no significant structure in the color-magnitude diagram.  A slight “wavy” structure remains in both the giant and dwarf residuals in the [M/H] range of $-$1.5 to $-$1.0, indicating that the adopted polynomial does not fully capture the detailed stellar behavior in this relatively higher-metallicity regime. However, its amplitude is much smaller than the 1$\sigma$ scatter of the residuals, and does not degrade photometric metallicity precision. The corresponding offset is also much smaller than the 1$\sigma$ scatter in $\Delta[\rm{M/H}]$, as shown in the middle panels of Figures\,\ref{fig:giant_result} and \ref{fig:dwarf_result}. Therefore, although the model could be further refined in this range, any gain in photometric metallicity precision would be marginal. 

The slight increase in scatter at the bright end is due to saturation of bright stars in the oMEGACat.II HST photometry, flagged as brightlist\_flag in that catalog.
\vskip 1cm

\begin{table*}
\centering
\caption{Fitting Coefficients for the F275W$-$F625W and F336W$-$F625W Colors in the Dwarf and Giant Samples}
\begin{tabular}{c|rr|rr}
\hline
\multirow{2}{*}{Coeff.} & \multicolumn{2}{c|}{Giants} & \multicolumn{2}{c}{Dwarfs} \\ 
                   & F275W$-$F435W                  & F336W$-$F435W                 & F275W$-$F435W & F336W$-$F435W \\
\hline
p0  & $-1.6172\times10^{-3}$ & $-8.4010\times10^{-4}$ & $6.8237\times10^{-3}$ & $1.6943\times10^{-3}$ \\
p1  & $8.3204\times10^{-3}$ & $4.4056\times10^{-3}$ & $-2.0717\times10^{-2}$ & $-5.5403\times10^{-3}$ \\
p2  & $-1.4615\times10^{-2}$ & $-8.3584\times10^{-3}$ & $2.1592\times10^{-2}$ & $6.2403\times10^{-3}$ \\
p3  & $1.1507\times10^{-2}$ & $7.0938\times10^{-3}$ & $-7.9454\times10^{-3}$ & $-2.4724\times10^{-3}$ \\
p4  & $-3.6526\times10^{-3}$ & $-1.9746\times10^{-3}$ & $2.3833\times10^{-5}$ & $-3.4490\times10^{-4}$ \\
p5  & $1.9887\times10^{0}$ & $4.3599\times10^{-1}$ & $-8.8507\times10^{0}$ & $-2.4340\times10^{0}$ \\
p6  & $-1.1663\times10^{1}$ & $8.2038\times10^{-1}$ & $2.0933\times10^{1}$ & $7.4513\times10^{0}$ \\
p7  & $2.2729\times10^{1}$ & $-8.4949\times10^{0}$ & $-2.4773\times10^{1}$ & $-8.3711\times10^{0}$ \\
p8  & $-1.5875\times10^{1}$ & $1.4666\times10^{1}$ & $9.3625\times10^{0}$ & $1.0799\times10^{0}$ \\
p9  & $2.6126\times10^{-3}$ & $-3.8900\times10^{-4}$ & $6.3191\times10^{-3}$ & $4.7825\times10^{-3}$ \\
p10 & $-1.9550\times10^{-2}$ & $1.2235\times10^{-2}$ & $-7.4405\times10^{-2}$ & $-2.0779\times10^{-2}$ \\
p11 & $1.2563\times10^{-1}$ & $-3.0423\times10^{-1}$ & $6.1580\times10^{-1}$ & $2.0508\times10^{-2}$ \\
p12 & $-7.0604\times10^{-1}$ & $2.3818\times10^{0}$ & $-2.0205\times10^{0}$ & $-9.3594\times10^{-2}$ \\
p13 & $-2.5365\times10^{-1}$ & $1.3311\times10^{0}$ & $2.5131\times10^{0}$ & $5.2600\times10^{-1}$ \\
p14 & $-3.6243\times10^{-2}$ & $2.5593\times10^{-1}$ & $-4.5326\times10^{-1}$ & $-3.0530\times10^{-1}$ \\
p15 & $5.7059\times10^{-1}$ & $-3.7598\times10^{0}$ & $4.1588\times10^{-1}$ & $1.4912\times10^{0}$ \\
p16 & $-2.8814\times10^{-5}$ & $-9.8715\times10^{-6}$ & $-4.3764\times10^{-5}$ & $-3.1452\times10^{-5}$ \\
p17 & $2.1939\times10^{-4}$ & $1.5541\times10^{-5}$ & $3.2837\times10^{-4}$ & $2.8009\times10^{-4}$ \\
p18 & $5.0464\times10^{0}$ & $-6.6228\times10^{0}$ & $1.6119\times10^{0}$ & $-3.9078\times10^{-2}$ \\
% ...
\hline
\end{tabular}

\label{tab:fit_coeff}
\end{table*}

\subsection{Test Results}\label{subsec:test}
As described in Section~\ref{sec:method}, we estimate photometric metallicities with a minimum-$\chi^2$ method. For each star, we evaluate $\chi^2$ on a [M/H] grid from $-3.0$ to $0.0$ in 0.01\,dex steps (301 points). The photometric metallicity, $\rm{[M/H]_{phot}}$, is the grid value at minimum $\chi^2$. We then compare $\rm{[M/H]_{phot}}$ to the spectroscopic metallicities, $\rm{[M/H]_{spec}}$, to evaluate our method.

\begin{figure*}
    \centering
    \includegraphics[width=\textwidth]{ 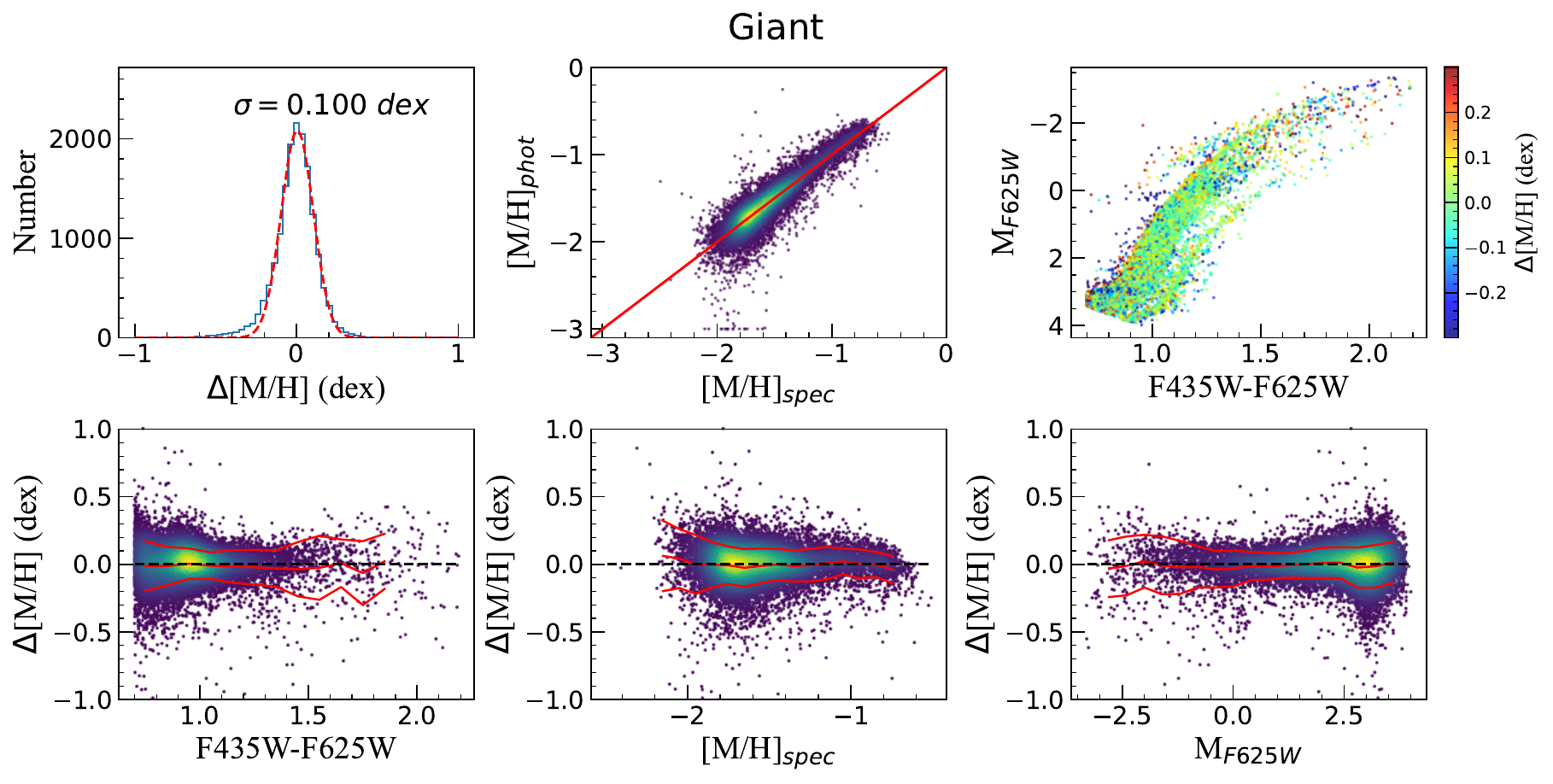}
    \caption{Photometric-metallicity results for giants. Top left: Histogram of $\Delta$[M/H] with a Gaussian fit (dashed-red line). Top middle: Comparison of photometric metallicities $\rm [M/H]_{phot}$ with MUSE spectroscopic metallicities $\rm [M/H]_{spec}$. Top right: $\Delta$[M/H] distribution in the color–magnitude diagram. Bottom panels: $\Delta$[M/H] versus F435W$-$F625W color, $\rm [M/H]_{spec}$, and $\rm M_{F625W}$, respectively. Points are color-coded by number density.
    }
    \label{fig:giant_result}
\end{figure*}

\begin{figure*}
    \centering
    \includegraphics[width=\textwidth]{ 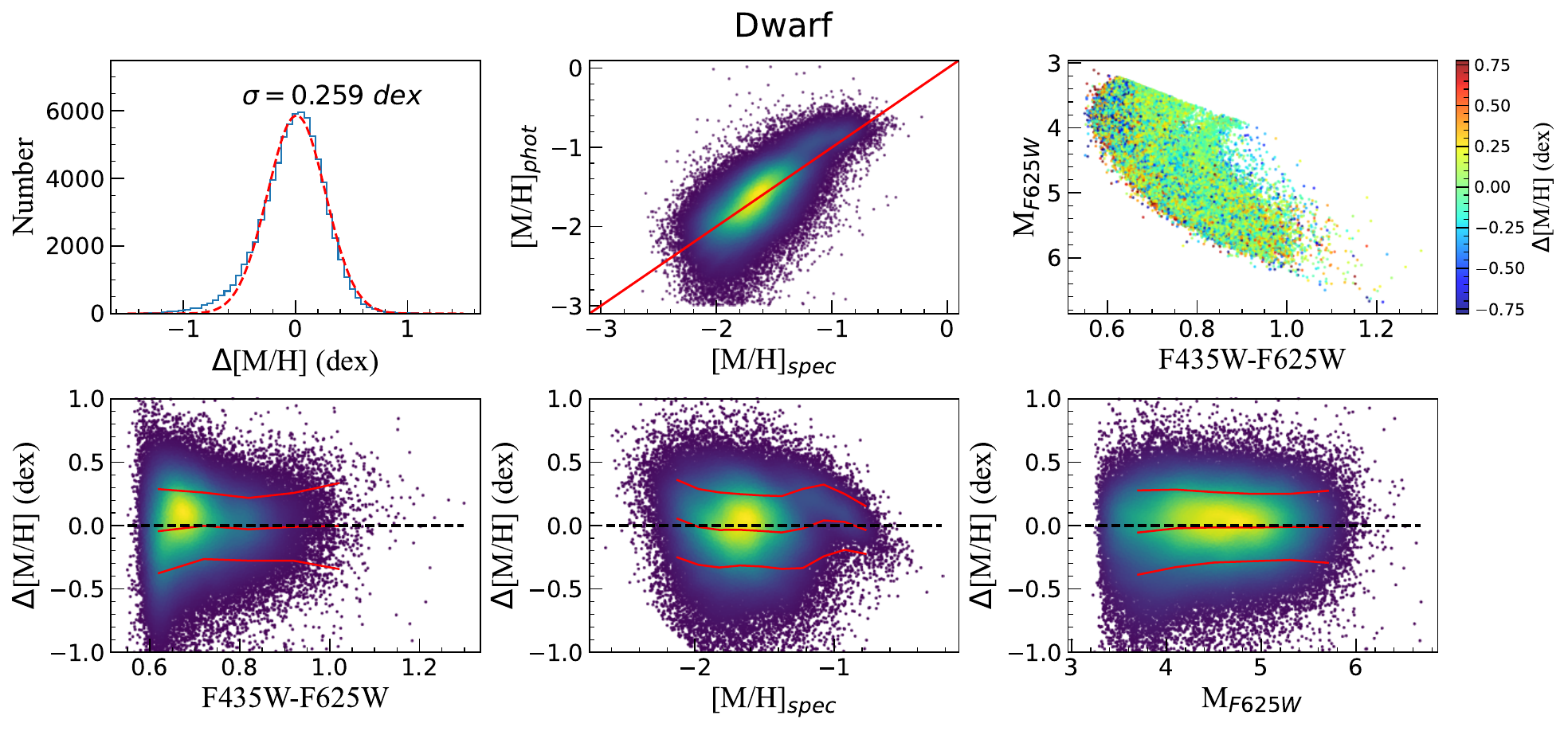}
    \caption{Same as Figure\,\ref{fig:giant_result}, but for dwarfs.}
    \label{fig:dwarf_result}
\end{figure*}

Figures\,\ref{fig:giant_result} and \ref{fig:dwarf_result} show the test results for giants and dwarfs, with residuals defined as $\Delta\rm{[M/H]} = {[M/H]_{phot}} - {[M/H]_{spec}}$. The mean $\Delta[\rm M/H]$ is close to 0.0\,dex for both samples. The dispersion in $\rm \Delta[M/H]$ is 0.10\,dex for giants and 0.26\,dex for dwarfs. Typical uncertainties in $\rm{[M/H]_{spec}}$ are 0.03\,dex (giants) and 0.13\,dex (dwarfs), reflecting the lower signal-to-noise ratios of the dwarf spectra. These imply intrinsic scatters of $\simeq 0.10$\,dex and $\simeq 0.22$\,dex in $\rm{[M/H]_{phot}}$ for giants and dwarfs, respectively. The residuals show no significant trends with F435W$-$F625W color, $\rm{[M/H]_{spec}}$, or $\rm M_{F625W}$, and no coherent structure in the color–magnitude diagram, indicating consistent model performance in the sample parameter space.

\section{Final catalog} \label{sec:fianl_cat}
We apply our models to the oMEGACat.II photometry to construct a photometric-metallicity catalog. We selected all sources with F336W, F435W, and F625W measurements, colors within the training-sample ranges ($\rm 0.7<F435W-F625W<2.2$ for giants, $\rm 0.55<F435W-F625W<1.3$ for dwarfs), and $\rm F625W<21$. To maximize the sample size, F275W photometry is optional. For sources without F275W, we derive metallicities from the F336W$-$F625W color using the same minimum-$\chi^2$ method. For sources with F275W, we follow Section~\ref{sec:method} and obtain $\rm[M/H]_{phot}$ from both F275W$-$F625W and F336W$-$F625W. This yields 367,324 stars (20,778 giants and 346,546 dwarfs), of which 249,016 stars have spectroscopic metallicities. For each source, the catalog lists oMEGACat.II photometry in all available filters and the derived photometric metallicity $\rm[M/H]_{phot}$ in column mh\_phot. The catalog will be publicly available at doi: \href{https://doi.org/10.12149/101818}{10.12149/101818}.

\begin{figure*}[htbp]
\centering
\subfigure{
\includegraphics[width=7cm]{ 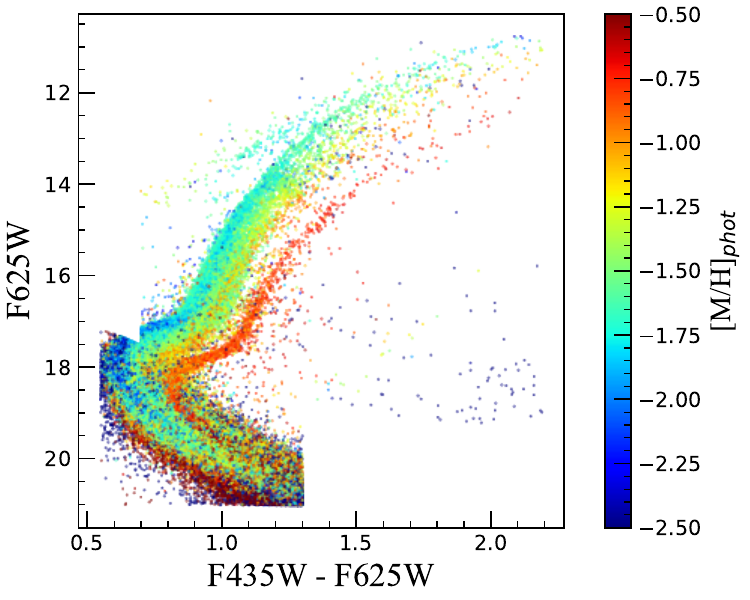}}
\subfigure{
\includegraphics[width=7cm]{ 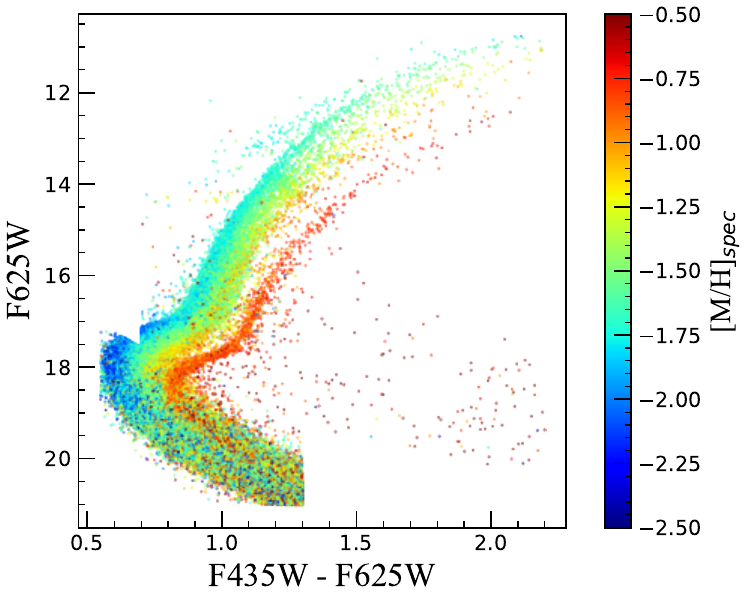}}
\caption{Color–magnitude diagram (CMD) color-coded by photometric metallicity [M/H]$_\mathrm{phot}$ (left) and spectroscopic metallicity $\rm [M/H]_{spec}$ (right).}
\label{fig:CMD_mh}
\end{figure*}

\begin{figure*}[htbp]
\centering
\subfigure{
\includegraphics[width=7cm]{ 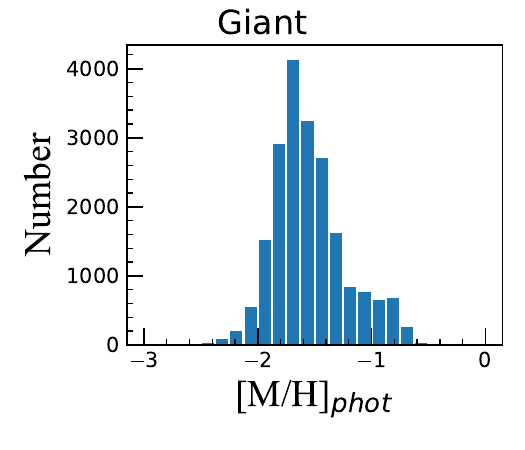}}
\subfigure{
\includegraphics[width=7cm]{ 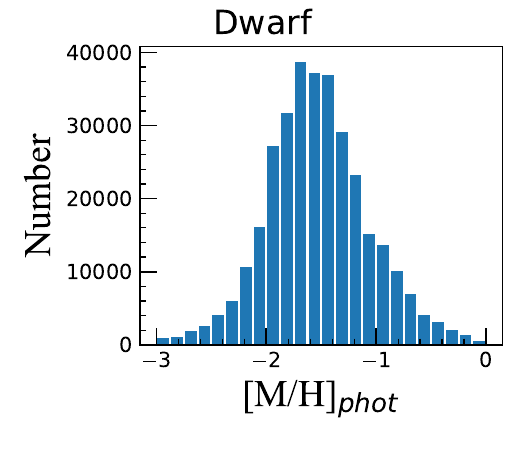}}
\caption{Distribution of photometric metallicity [M/H]$_\mathrm{phot}$ for giants (left) and dwarfs (right). }
\label{fig:hist_mh}
\end{figure*}

Figure\,\ref{fig:CMD_mh} shows color–magnitude diagrams color-coded by $\rm [M/H]_{phot}$ and $\rm [M/H]_{spec}$. The subpopulations are similarly distinguishable with both photometric and spectroscopic metallicities, supporting the reliability of our photometric results. Figure\,\ref{fig:hist_mh} presents the $\rm [M/H]_{phot}$ distributions for giants and dwarfs separately. We apply a boundary cut, discarding 245 giants and 18,674 dwarfs.
The mean $\rm [M/H]_{phot}$ is $-1.53$ for dwarfs and $-1.55$ for giants, matching the MUSE spectroscopic metallicities in 
\citeauthor{omegacat3} (\citeyear{omegacat3}, hereafter oMEGACat.III). As in the spectroscopic data, giants mainly lie between $-2.0$ and $-0.5$ and show multiple peaks. 

\begin{figure*}
    \centering
    \includegraphics[width=14cm]{ 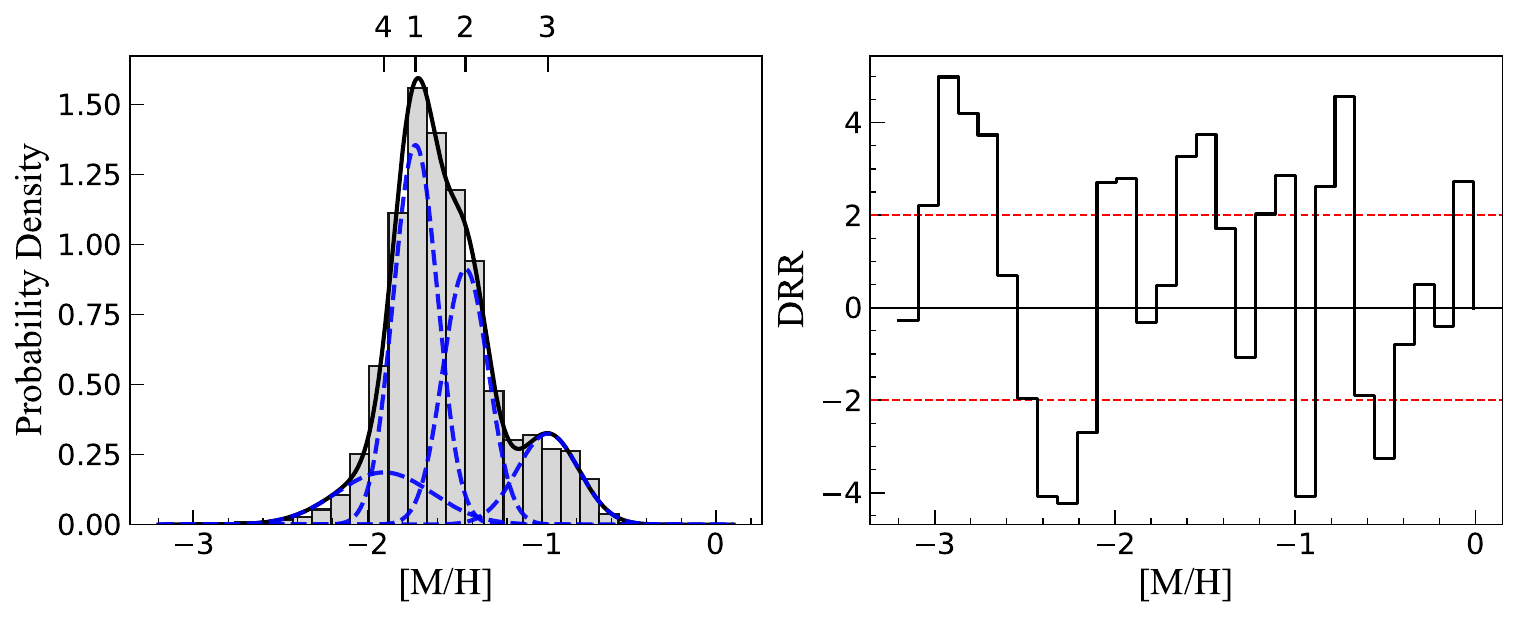}
    \caption{Left panel: Probability density distribution of [M/H] for the giants $\rm [M/H]_{phot}$, fitted with a Gaussian mixture model, shown as a black solid line, with each component (4 in total) shown as a blue dashed line.
    Right panel: The Double Root Residual (DRR) for GMM fits (see text for description). The dot-dashed lines at $\pm 2$ indicate an approximate 95\% significance level.
    }
    \label{fig:GMM}
\end{figure*}

We modeled the metallicity distributions of giants with a Gaussian mixture model (GMM), shown in the left panel of Figure\,\ref{fig:GMM} and Table\,\ref{tab:gmm_components}. Using the Bayesian information criterion (BIC), we find that four Gaussians provide the optimal description, with additional components yielding only minor improvements. 
To evaluate the fit, we computed Double Root Residuals (DRRs) following \cite{Carollo_2010}; the DRR distribution is shown in the right panel of Figure\,\ref{fig:GMM}.\footnote{The DRR plot is a variation of a hanging histogram, which is a (non-normalized) difference plot between binned data and a proposed fitting function. One limitation of the hanging histogram is that it gives equal weight to residuals from bins near the extrema of a distribution (which are usually poorly populated) as to residuals near the middle of a distribution (which are usually well-populated, and therefore subject to greater $\sqrt{N}/N$ variations). A square-root transformation applied to the residuals provides a desirable variance stabilization.  It also graphically emphasizes \textit{where} in a distribution a lack of fit between data and model exists -- the square-root transformation puts the residuals throughout the fit on an equal footing.} While the four-component GMM reproduces the overall metallicity distribution reasonably well, the DRRs reveal coherent residuals at several metallicities, reaching $\sim 2\sigma$. Increasing the number of components does not substantially improve these residuals, suggesting that the MDF of $\omega$~Cen is intrinsically complex and likely reflects a prolonged, chemically inhomogeneous enrichment history that cannot be fully captured by a small set of discrete Gaussians. 

In oMEGACat.III, the authors used a Gaussian mixture model on MUSE-based metallicities for 11,050 stars and identified 11 Gaussian components. Although the BIC favored 11 components, they noted that models with 8 or even 5 components fit nearly as well because several components have very small amplitudes or strongly overlap within the main metallicity peak. The four peaks in our photometric metallicity distribution agree with the main peaks in oMEGACat.III, and our sample—about twice as large as the spectroscopic one—offers enough statistical power to robustly characterize the principal stages of chemical enrichment.

\begin{table}
\centering
\caption{Gaussian Components of the Metallicity Distribution}
\label{tab:gmm_components}
\begin{tabular}{cccc}
\hline\hline
Index  &
Mean &
Intrinsic Standard Deviation &
Fraction \\
\# &  & (dex) & (\%) \\
\hline
1 & $-1.725$ & $0.126$ & $42.9$ \\
2 & $-1.438$ & $0.131$ & $29.8$ \\
3 & $-0.965$ & $0.177$ & $14.4$ \\
4 & $-1.905$ & $0.275$ & $12.9$ \\
\hline
\end{tabular}
\tablecomments{
The Gaussian components are ordered by decreasing fraction of stars.}
\end{table}

\vskip 4 cm
\section{Two-Dimensional Spatial Distribution and Radial Gradients of Metallicity} \label{sec:metal_dist}

The 20,533 giants in our final catalog  after a boundary cut form a more complete data set than the MUSE spectroscopic giant sample of oMEGACat.III, and are used to study the spatial distribution of photometric metallicities. We apply a procedure similar to that in oMEGACat.III.

\begin{figure*}
    \centering
    \subfigure{
    \includegraphics[width=10cm]{ 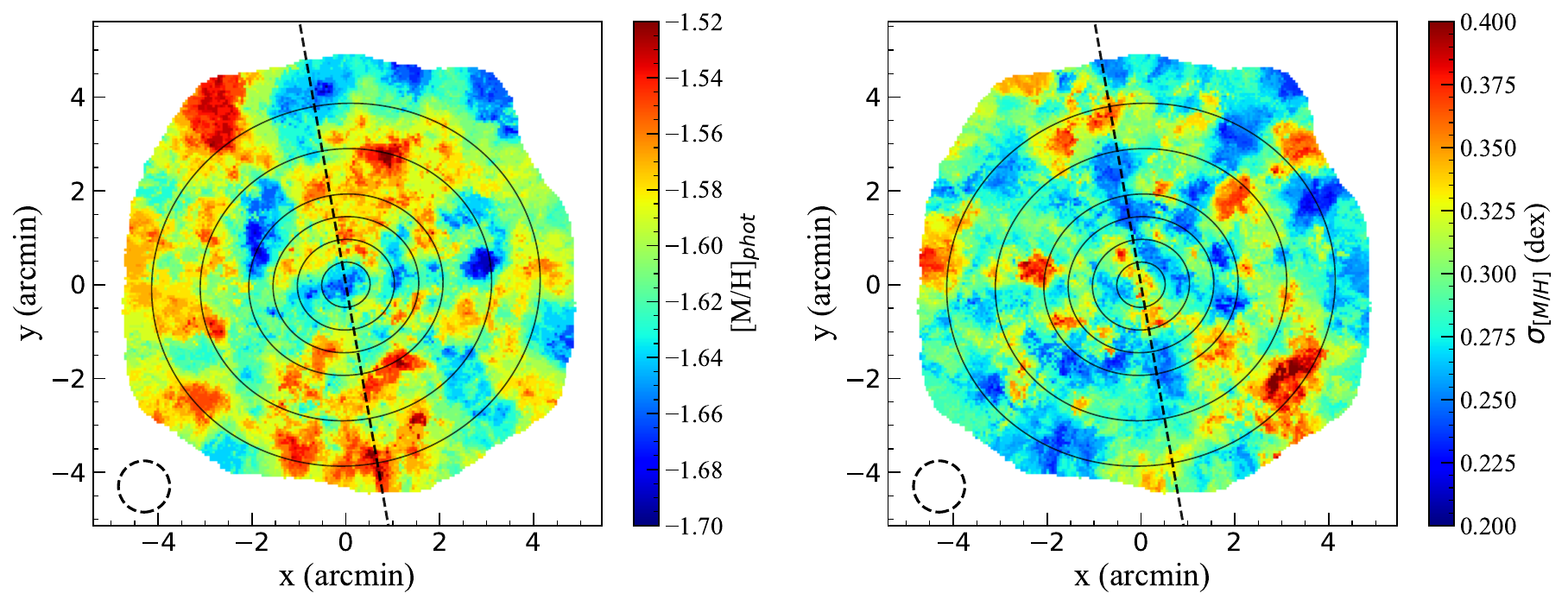}}
    \subfigure{
    \includegraphics[width=10cm]{ 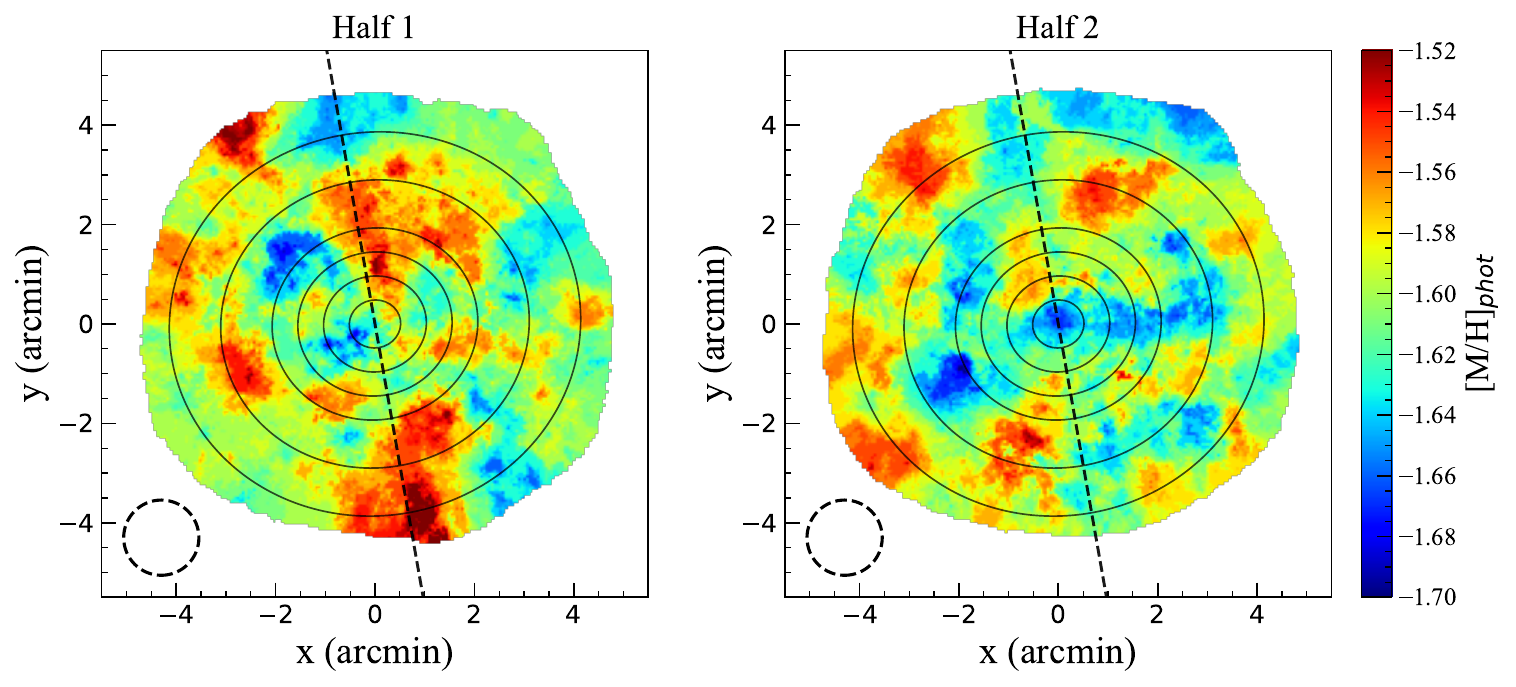}}
    \subfigure{\includegraphics[width=5cm]{ 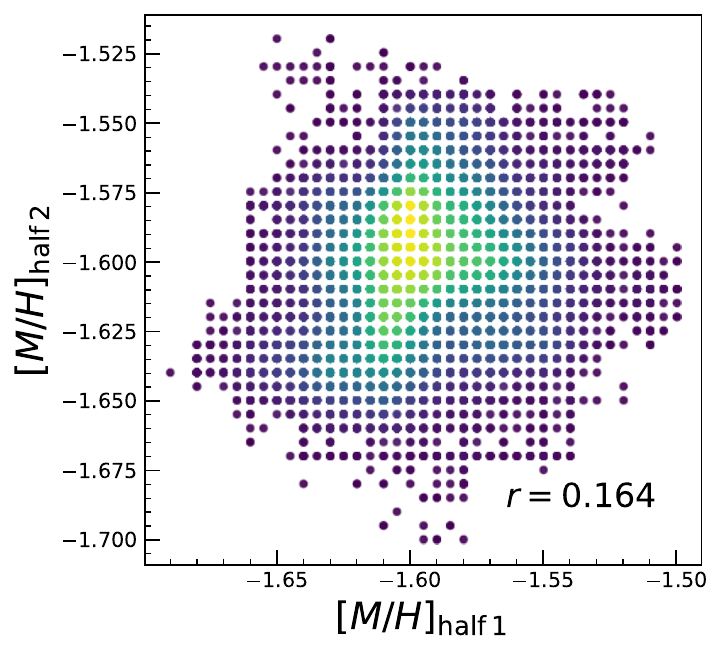}}
    \caption{The top two panels show the median $\rm [M/H]_{phot}$ map (left) and its dispersion $\sigma_{\rm [M/H]}$ (right), both derived from the 200 nearest neighbors. The dashed-black circle in the top-left panel indicates the mean radius of these neighbors. Black elliptical annuli mark the bin edges used in Section\,\ref{sec:metal_dist}, and their minor axis, aligned with the rotation axis, is shown as a dashed-black line. The bottom three panels show the median $\rm [M/H]_{phot}$ for Half1, for Half2, and the correlation between $[\rm M/H]_{Half1}$ and $[\rm M/H]_{Half2}$, with correlation coefficient $r=0.164$.
    }
    
    \label{fig:mh_dist}
\end{figure*}

We construct smooth maps of the median $\rm [M/H]_{phot}$ for each position using its 200 nearest neighbors (top-left panel of Figure\,\ref{fig:mh_dist}); the corresponding local dispersion, defined as half the 68\% range from the same neighbor sets, is shown in the top-right panel. The mean radius of the 200 neighbors is 0.55\,arcmin, indicated by the dashed-black circle in the top panels.

oMEGACat.III reported a ring-like two-dimensional metallicity distribution, which they attributed to intermediate-[M/H] populations possibly accreted later and not yet well mixed. In contrast, our median $\rm [M/H]_{phot}$ map does not show a clear ring, but instead displays an apparently irregular spatial pattern. 

To test the robustness of this pattern, we randomly split the giant sample into two equal subsamples. The median $\rm [M/H]_{phot}$ maps of these half-samples and their correlation are shown in the bottom panels. The correlation is weak (coefficient 0.164), and repeated random splits yield similarly low values. We also build a control sample by assigning the observed metallicity distribution to random $(x,y)$ positions; its median map shows comparable spatial fluctuations. These analyses indicate that the observed irregular structural pattern is unlikely to have substantial physical significance.

\begin{figure*}
\centering
\includegraphics[width=\textwidth]{ 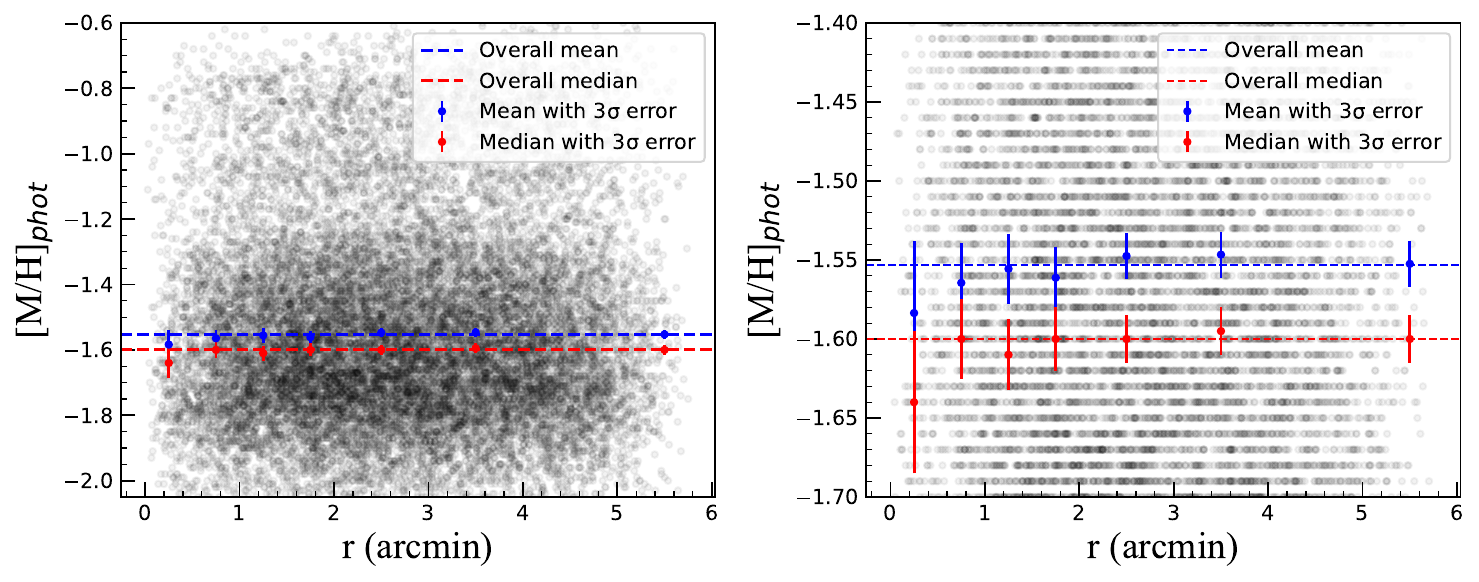}
\caption{The left panel shows the full [M/H] range, and the right panel zooms in on metallicities near the mean and median. Black dots mark all stars in the analysis, red points show the median in each bin, and blue points show the mean. Dashed lines indicate the overall median and mean of the full sample.}
\label{fig:mh_vs_R}
\end{figure*}

We also used the same elliptical bins as in oMEGACat.III to analyze metallicity gradients, given the elongation of the cluster in the sky. The position angle is $100^\circ$, the ellipticity is 0.07, and the bin equivalent radii are (0.5, 1, 1.5, 2, 3, 4, 7) arcmin. These bin edges are shown in black in the top panels of Figure\,\ref{fig:mh_dist}. In all subsequent analyses we compute the median circular radius in each elliptical bin and plot that value. Figure\,\ref{fig:mh_vs_R} shows the mean and median metallicity for each bin, with a 99.7\% ($3\sigma$) error bars derived via bootstrapping.
We detected no significant gradient within the cluster’s half-light radius, and the gradients are notably flatter than those reported in oMEGACat.III. In particular, unlike oMEGACat.III, we see no $3\sigma$ excesses above the global median $\rm [M/H]_{phot}$ in bins 2--4. 

Thus, we find no strong chemical evidence for merging within the half-light radius of $\omega$~Centauri. Although this does not exclude a merger-driven origin, our results are broadly consistent with $\omega$~Centauri being the remnant nucleus of a dwarf galaxy accreted and tidally stripped by the Milky Way in the distant past.


\section{Summary and Perspectives} \label{sec:summary}
$\omega$~Centauri is the most massive and chemically complex multi-population globular cluster, exhibiting a wide metallicity range from [Fe/H]~$\approx -2.2$ to $-0.5$. It has long been hypothesized to be the tidally stripped remnant nucleus of a dwarf galaxy that merged with the Milky Way in the distant past. Extensive photometric and spectroscopic observations have been carried out over the years to investigate its complex stellar population structure.

In this study, we employ a metallicity- and $\rm M_{F625W}$-dependent stellar-locus technique to derive photometric metallicities for stars in $\omega$~Cen, using \textit{HST} photometry in combination with MUSE spectroscopic data. Validation tests indicate that the typical metallicity precision is 0.10\,dex for giant stars, and 0.22\,dex for dwarf stars.

We construct a photometric metallicity catalog based on simultaneous F336W, F435W, and F625W observations (and F275W where available), comprising 367,324 stars (20,778 giants and 346,546 dwarfs). From this sample, 20,533 giants with a boundary cut are selected to investigate the spatial metallicity distribution and possible gradients. 
Our analysis reveals no significant metallicity gradient within the half-light radius of the cluster, consistent with the findings of \citealt{omegacat3}. Moreover, the ring-like structure reported there appears weaker in our data. 
We consider no physically meaningful irregularities in the metallicity structure, revealing the subpopulations seem to be well mixed inside the half-light radius.

The catalog presented here significantly expands the metallicity sample available for $\omega$~Cen, and the methodology developed in this work can be  applied to other \textit{HST} datasets to estimate photometric metallicities.

\begin{acknowledgments}

This work is supported by the National Natural Science Foundation of China through the project NSFC 12222301, 12173007, 124B2055,
the National Key Basic R\&D Program of China via 2024YFA1611901 and
2024YFA1611601.  
This work is also supported by the China Manned Space Program
with grant no. CMS-CSST-2025-A13.
T.C.B. acknowledges partial support from grants PHY 14-30152; Physics Frontier Center/JINA Center for the Evolution of the Elements (JINA-CEE), OISE-1927130: The International Research Network for Nuclear Astrophysics (IReNA), awarded by the US National Science Foundation, and DE-SC0023128; the Center for Nuclear Astrophysics Across Messengers (CeNAM), awarded by the U.S. Department of Energy, Office of Science, Office of Nuclear Physics.  His participation in this work was initiated
by conversations that took place during a visit to China in 2019, supported by a PIFI Distinguished Scientist award from the Chinese Academy of Science.

\end{acknowledgments}

\bibliographystyle{aasjournal}
\bibliography{main}

\end{document}